\documentclass[aps,pra,twocolumn,showpacs,superscriptaddress]{revtex4}
\usepackage{graphicx,bm,dcolumn,amsmath,amssymb,amsfonts}
\newcommand{\vsigma}{\bm{\sigma}}
\newcommand{\vkappa}{\bm{\kappa}}
\newcommand{\vn}{\bm{n}}
\newcommand{\vm}{\bm{m}}
\newcommand{\vk}{\bm{k}}
\begin{document}


\title{Implementation of holonomic quantum gates by an isospectral deformation \\of an Ising dimer chain} 

\author{Yukihiro Ota}
\affiliation{
Research Center for Quantum Computing, Interdisciplinary Graduate School of Science and Engineering, Kinki University, 
3-4-1 Kowakae, Higashi-Osaka, 577-8502, Japan}
\author{Masamitsu Bando}
\altaffiliation[Current address:]{
Department of Physics, Graduate School of Science, Osaka University, 1-1
Machikaneyama, Toyonaka, Osaka 560-0043, Japan 
}
\affiliation{
Department of Physics, Kinki University, 
3-4-1 Kowakae, Higashi-Osaka, 577-8502, Japan}
\author{Yasusi Kondo}
\affiliation{
Research Center for Quantum Computing, Interdisciplinary Graduate School of Science and Engineering, Kinki University, 
3-4-1 Kowakae, Higashi-Osaka, 577-8502, Japan}
\affiliation{
Department of Physics, Kinki University, 
3-4-1 Kowakae, Higashi-Osaka, 577-8502, Japan}
\author{Mikio Nakahara}
\affiliation{
Research Center for Quantum Computing, Interdisciplinary Graduate School of Science and Engineering, Kinki University, 
3-4-1 Kowakae, Higashi-Osaka, 577-8502, Japan}
\affiliation{
Department of Physics, Kinki University, 
3-4-1 Kowakae, Higashi-Osaka, 577-8502, Japan}
\date{\today}

\begin{abstract}
We exactly construct one- and two-qubit holonomic quantum gates
in terms of isospectral deformations of an Ising model Hamiltonian.
A single logical qubit is constructed out of two spin-$\frac{1}{2}$ particles;
the qubit is a dimer. We find that the holonomic gates obtained are discrete
but dense in the unitary group.
Therefore an approximate gate for a desired one can be constructed with
arbitrary accuracy.
\end{abstract}


\maketitle

\section{Introduction}
\label{sec:intro}
A reliable implementation of a quantum gate is required to realize
quantum computing. 
A quantum gate is often realized by manipulating the parameters 
in the Hamiltonian of a
system so that the time-evolution operator results in a desired
unitary gate. 
On the other hand, when the system has a degenerate energy eigenvalue,
adiabatic parameter control allows us to construct a quantum gate
employing non-Abelian holonomy\,\cite{WilczekZee1984}. 
Holonomy corresponds to the difference between the initial and the
final quantum states under an adiabatic change of parameters
along a closed path (loop) in the parameter manifold $\mathcal{M}$ \cite{Nakahara2003}. 
Therefore, a desired quantum gate can be implemented by choosing a
proper closed loop in $\mathcal{M}$. 
This scheme is called the holonomic quantum computing (HQC). 
The idea was suggested first in Ref. \cite{ZanardiRasetti1999}
and has been developed subsequently by many authors \cite{Fujii2000,Fujii2001,
NiskanenNakaharaSalomaa2003,TanimuraHayashiNakahara2004,
KarimipourMajd2004,TanimuraNakaharaHayashi2005}.
Holonomy is geometrical by nature, and hence it is independent of how
fast the loop in the parameter manifold is traversed.
In addition, if the lowest eigenspace of the spectra is employed as a
computational subspace, it is free from errors caused by spontaneous decay. 
Thus, HQC is expected to be robust against noise and
decoherence \cite{SolinasZanardiZanghi2004}.

In spite of its mathematical beauty, physical implementation of HQC
is far from trivial.
The difficulties are (i) to find a quantum system in which
the lowest energy eigenvalue is degenerate and (ii) to design
a control which leaves the ground state
degenerate as the loop is traversed. 
Several theoretical ideas have been proposed in
linear optics~\cite{PachosChountasis2000},
trapped ions~\cite{DuanCiracZoller2001,RacatiCalarcoZanardiCiracZoller2002,Pachos2002}, and Josephson junction qubits~\cite{FaoroSiewertFazio2003}. 
Recently, an experiment following the proposals made in
\cite{DuanCiracZoller2001,RacatiCalarcoZanardiCiracZoller2002},
where the coding space is not the lowest eigenspace, has been
reported~\cite{GotoIchimura2007}. 

Karimipour and Majd\,\cite{KarimipourMajd2005} proposed HQC
with a spin chain model.
A single logical qubit is represented by two spin-$\frac{1}{2}$
physical spins; the qubit is a dimer.
In the dimer, the spins interact with each other
through a Heisenberg-type interaction and its lowest eigenvalue is
doubly degenerate for a particular set of the coupling constant
and the Zeeman energies. 
The spin chain model may be applicable to various physical systems,
such as solid state systems.  
However, there is room to reconsider and improve their proposal.
In particular, we want to investigate whether the Heisenberg-type
interaction is essential for HQC in a spin chain model.   
Such consideration is necessary for clarifying the applicability of
their proposal. 
In this paper, we closely follow the discussion given
in Ref.~\cite{KarimipourMajd2005} and construct holonomic quantum
gates using isospectral deformations of an Ising model, instead of
a Heisenberg model.
In addition, we explicitly require that a path in the parameter manifold
${\mathcal M}$ be closed for a holonomy to be well-defined.

The paper is organized as follows.
We briefly review holonomy associated with adiabatic time evolution of a
quantum system in Sec.~\ref{sec:rev_holonomy}.
We show how to construct one- and two-qubit gates as 
holonomies in Sections \ref{sec:hqg1} and \ref{sec:hqg2}.
Section \ref{sec:summary} is devoted to summary.

\section{Holonomy}
\label{sec:rev_holonomy}
Let us consider a Hamiltonian $H$ acting on a Hilbert space
$\mathcal{H}$ (${\rm dim}\,\mathcal{H}=N$), whose
$l$th eigenvalue is denoted as $E_{l}$. 
In particular, $l=0$ refers to the lowest eigenvalue. 
The $l$th eigenvalue is $g_l$-fold degenerate and its
eigenvectors are written as $|l,\,i \rangle$ ($i=1,\,2,\ldots,g_{l}$). 
We assume $\langle l,\,i|m,\,j \rangle = \delta_{lm} \delta_{ij}$.
Note that $N = \sum_{l} g_{l}$.

An isospectral deformation of $H$ is accomplished by 
\begin{equation}
 H(\tau) = g(\tau)\,H\,g^{\dagger}(\tau)
\quad
(0 \le \tau \le 1), 
\label{eq:gen_iso_def}
\end{equation}
where $g(\tau)$ $\in {\rm U}(N)$. As a result, no level crossing takes place
during the Hamiltonian deformation. We normalize $\tau \in [0, 1]$ so that
$H(0) = H(1) = H$. This condition implies that the curve in $\mathcal{M}$
be closed.
The symbol $\tau$ in Eq.~(\ref{eq:gen_iso_def}) is the normalized dimensionless
time: $\tau={t}/{T}$ ($0\le t\le T$), where $T$ is the total time to
traverse the loop.
Note that $T$ is long enough so that the adiabatic approximation may be
justified~\cite{adtheorem}. 

We consider the particular isospectral deformation 
\begin{eqnarray}
H(\tau) = e^{X\tau} H e^{-X\tau}, 
\label{eq:isodeform}
\end{eqnarray}
following~\cite{KarimipourMajd2004,KarimipourMajd2005},
where $X$ is a constant anti-Hermitian matrix.
We require
\(
[H,\,X]\neq 0
\) and
\begin{equation}
 e^{X} = \openone,
\label{eq:cond_colsed}
\end{equation}
where $\openone$ is the unit matrix. The first condition is required
to implement non-trivial gates while
Eq.~(\ref{eq:cond_colsed}) ensures that $H(1) = H(0) =H$.

We readily obtain the instantaneous eigenvalues $ E_{l}(\tau)$ and
eigenvectors $ |l, i; \tau \rangle$ of $H(\tau)$ as follows
\cite{KarimipourMajd2004,KarimipourMajd2005}: 
\begin{eqnarray}
 E_{l}(\tau) =E_{l}, 
\quad 
 |l,\,i;\,\tau\rangle = e^{X\tau}|l,\,i\rangle,
\label{eq:solution_isodeig}
\end{eqnarray}
which implies that no level crossing occurs during the deformation: 
$E_{l}(\tau) \neq E_{l^{\prime}}(\tau)$ 
($l\neq l^{\prime}, \forall \tau \in [0, 1]$). 
We will exclusively work with the
ground state ($l=0$) from now on and drop the index $l=0$ whenever
it cause no confusion. 

We use the unit in which $\hbar=1$ through the paper. 
According to the adiabatic theorem, when the initial state
$|\psi(\tau=0)\rangle$ is in the lowest eigenspace, the final state
$|\psi(\tau=1)\rangle$ remains within this subspace. 
Then, we find
\begin{eqnarray}
 |\psi(\tau=1)\rangle
=
 e^{-iE_{0}T} \,\Gamma\,|\psi(\tau=0)\rangle,
\label{eq:solution}
\end{eqnarray}
where $ e^{-iE_{0}T}$ is the dynamical phase and $\Gamma \in {\rm U}(g_{0})$.
Equation (\ref{eq:solution}) can be regarded as a gate operation
connecting the initial and final states. 
The unitary matrix $\Gamma$ in Eq.~(\ref{eq:solution}) is the holonomy
associated with the cyclic deformation Eq.~(\ref{eq:isodeform}) of the
Hamiltonian and we write it as
\begin{eqnarray}
 \Gamma &=& e^{-A}. 
\label{eq:holonomy_isod}
\end{eqnarray}
The anti-Hermitian connection $A$ in Eq.~(\ref{eq:holonomy_isod}) is
given, in terms of $X$, as 
\begin{eqnarray}
 A &=& \sum_{i,\,j=1}^{g_{0}}\langle i|X|j\rangle
|i\rangle\langle j|.
\label{eq:aconnection_isod}
\end{eqnarray}
In order to calculate $A$, the instantaneous eigenvector
$|0,\,i;\,\tau\rangle$ in Eq.~(\ref{eq:solution_isodeig}) is substituted
into  
\begin{eqnarray}
 A_{ij}(\tau)
=
\langle i;\,\tau| \frac{d}{d\tau}|j;\,\tau\rangle,
\label{eq:element_cnnt_isod}
\end{eqnarray} 
which is the definition of the $ij$ component of the connection $A$.

\section{One-qubit gates}
\label{sec:hqg1}
\subsection{Hamiltonian}
\label{subsec:hamiltonian1}
\begin{figure}[htbp]
\centering
 \scalebox{0.4}[0.4]{\includegraphics{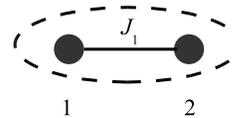}}
\caption{
\label{fig:onedimer}
Dimer consists of two spin-$\frac{1}{2}$ particles.
}
\end{figure}

We introduce a Hamiltonian
\begin{equation}
 H_{\rm 1D}
= -\omega \sigma_{1z} - \omega\sigma_{2z}
+ J_1 \sigma_{1z}\sigma_{2z}
\label{eq:Hamiltonian_for_single}
\end{equation}
as $H$ in Eq.~(\ref{eq:isodeform}),
where $\sigma_{ka}$ is the $a$-component of Pauli matrices
of the $k$th spin ($k=1,2$ and $a=x,y,z$).  
Equation (\ref{eq:Hamiltonian_for_single}) corresponds to
the Hamiltonian of two homogeneous spin-$\frac{1}{2}$ particles interacting
with each other via an Ising-type interaction
as depicted in Fig.~\ref{fig:onedimer}.
The strength of the interaction is parameterized by $J_1$,
while that of the field by $\omega$.
We assume $\omega>0$ and $J_1>0$ without loss of generality.

We denote the eigenvectors of $\sigma_{z}$ as follows:
\(
\sigma_{z}|+\rangle = |+\rangle
\) and
\(
\sigma_{z}|-\rangle = -|-\rangle
\).
Furthermore, we introduce the following vectors which are 
eigenvectors of $H_{\rm 1D}$: 
\(
 |T_{+}\rangle = |++\rangle
\), 
\(
 |T_{0}\rangle = \frac{1}{\sqrt{2}}(|+-\rangle +|-+\rangle)
\), 
\(
 |T_{-}\rangle = |--\rangle 
\), and 
\(
 |S_{0}\rangle = \frac{1}{\sqrt{2}}(|+-\rangle - |-+\rangle)
\), 
where $|++ \rangle$ denotes
$|+ \rangle \otimes |+ \rangle$, and so on. 
The eigenvalues of $|T_{+}\rangle, |T_{0}\rangle, |T_{-}\rangle$, and 
$|S_{0}\rangle$ are
$-2\omega +J_1, -J_1, 2\omega +J_1$ and $-J_1$, respectively. 

In case $\omega=J_1$, there exists $3$-fold degenerate
lowest energy eigenvalue $-J_1$.
The lowest eigenspace is spanned by the three eigenvectors
\(
|T_{+}\rangle
\),
\(
|T_{0}\rangle
\) and
\(
|S_{0}\rangle
\).
The unique excited state is $|T_{-}\rangle$ and
the energy difference between $|T_{-}\rangle$ and the ground state
is $4J_1$.

In this paper, we choose
\(
 |0\rangle_{\rm L} = |T_{+}\rangle
\) and 
\(
 |1\rangle_{\rm L} = |T_{0}\rangle
\) as basis vectors of the logical qubit. 
Namely, the coding space for a single qubit is 
\(
 \mathcal{C}_{1}={\rm Span}\{|T_{+}\rangle,\,|T_{0}\rangle\}
\). 
We note here that a qutrit may
be implemented by using the three ground state eigenvectors.
Although this coding is potentially interesting, it is
beyond the scope of this paper.

\subsection{Implementation of one-qubit gates}
\label{subsec:1qg}
One-qubit gates are implemented by choosing $X$ in Eq.~(\ref{eq:isodeform})
as\,\cite{KarimipourMajd2005}
\begin{equation}
 X =
i \vn\Omega\cdot (\vsigma_{1} + \vsigma_{2}),
\label{eq:X_1D}
\end{equation}
where $\vn$ is a unit vector in $\mathbb{R}^{3}$, while
$\Omega$ is a positive real number. 
It should be emphasized that undesired transitions into the
irrelevant subspace ${\rm Span}\{|S_{0}\rangle\}$
do not occur for this choice of $X$. This is easily seen from 
the identity $(\sigma_{1a} + \sigma_{2a})|S_0 \rangle = 0$. 
We obtain the anti-Hermitian connection restricted
to the coding space $ \mathcal{C}_{1}$ as follows: 
\begin{eqnarray}
 \left. A \right|_{\mathcal{C}_{1}}
=
 i\Omega [
n_{z}(I_{\rm L} + \sigma_{{\rm L}z}) 
+ \sqrt{2}(n_{x}\sigma_{{\rm L}x}
+ n_{y}\sigma_{{\rm L}y})
].\label{eq:aconnection_code}
\end{eqnarray}
where 
\(
 \sigma_{{\rm L}x}
= |T_{+}\rangle\langle T_{0}|
+ |T_{0}\rangle\langle T_{+}|
\), 
\(
 \sigma_{{\rm L}y}
= -i(|T_{+}\rangle\langle T_{0}|
- |T_{0}\rangle\langle T_{+}|)
\), 
\(
 \sigma_{{\rm L}z}
= |T_{+}\rangle\langle T_{+}|
- |T_{0}\rangle\langle T_{0}|
\) and 
\(
 I_{\rm L}
= |T_{+}\rangle\langle T_{+}|
+ |T_{0}\rangle\langle T_{0}|
\). 
Note that the presence of a closed loop in $\mathcal{M}$ is essential to
obtain Eq.~(\ref{eq:aconnection_code}). 

Using Eqs.~(\ref{eq:holonomy_isod}) and (\ref{eq:aconnection_code}), we
find the unitary operator 
\(
\Gamma 
= e^{-i\Omega n_{z}}
e^{-i\Omega [n_{z}\sigma_{{\rm L}z}
+ \sqrt{2}(n_{x}\sigma_{{\rm L}x}
+ n_{y}\sigma_{{\rm L}y})
]}
\). 
The condition $H_{\rm 1D}(1) = H_{\rm 1D}(0)$ is necessary for the
closure of a loop in $\mathcal{M}$\,\cite{comment}.  
As for Eq.~(\ref{eq:X_1D}), this condition is obviously
satisfied by taking
$\Omega = \kappa \pi $ ($\kappa\in\mathbb{N}$), where 
$\kappa$ is interpreted as the winding number of
a loop in $\mathcal{M}$. 

Summarizing the above arguments, we find that $X$ given in
Eq.~(\ref{eq:X_1D}) implements a single qubit holonomic gate
\begin{eqnarray}
 \Gamma^{(1)}_{\vm}(\kappa)
&=&
 e^
{-i\kappa \pi n_{z}}\,
e^{-i\theta_{\kappa}\vm \,\cdot\vsigma_{\rm L}}.
\label{eq:shqg}
\end{eqnarray}
where 
\(
 \vm
=
\frac{1}{\sqrt{2-n_{z}^{2}}}
\left(
\sqrt{2}n_{x},\,\sqrt{2}n_{y},\,n_{z}
\right)
\) and 
\(
\theta_{\kappa}=\kappa\pi\sqrt{2-n_{z}^{2}}
\). 
In order that the holonomic quantum gate is non-trivial, the condition 
$[H_{\rm 1D},\,X] \ne 0$ must be satisfied. 
This condition is equivalent to $|n_{z}| \neq 1$. 

Note that the set of the rotation angles $\theta_{\kappa}$ is discrete
when $n_z$ is fixed, and hence it is impossible to vary
$\theta_{\kappa}$ continuously in Eq.~(\ref{eq:shqg}). 
Nevertheless, we can find a quantum gate that approximates the desired
one with arbitrary accuracy 
provided that $\sqrt{2-n^{2}_{z}}$ is an irrational number~\cite{comment2}. 

\subsection{Examples}
\subsubsection{Hadamard gate}
\label{subsec:Hadamard}
\begin{figure}[htbp]
\centering
 \scalebox{1.0}[1.0]{\includegraphics{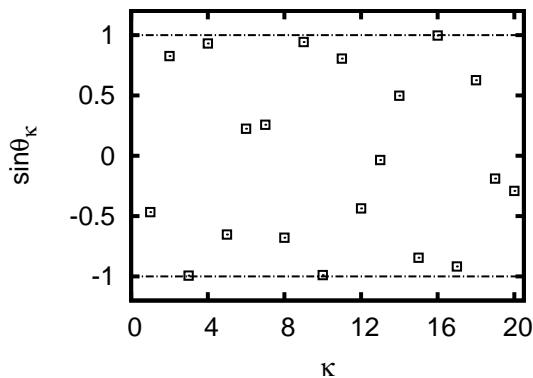}}
\caption{\label{fig:hadamard} 
$\sin\theta_{\kappa}$ as a function of $\kappa$. 
}
\end{figure}

The Hadamard gate, up to an irrelevant overall phase,
is implemented by taking both $|\sin \theta_{\kappa}| =1$ 
and $\vm =(1 ,0 ,1)/\sqrt{2}$ in
$\Gamma^{(1)}_{\vm}(\kappa)$.
Note that $n_z= \sqrt{2/3}$ and
$\theta_{\kappa} = {2\kappa\pi}/{\sqrt{3}}$ 
for the choice of $\vm$ and $|\sin \theta_{\kappa}| = 1$ cannot be
satisfied exactly for any $\kappa$. 
However we show that the Hadamard gate may be implemented
with good accuracy for some $\kappa$.
Figure \ref{fig:hadamard} shows $\sin \theta_{\kappa}$ as a function of 
$\kappa$, from which we find that
$|\sin \theta_{\kappa}| \simeq 1$ for $\kappa=3,10$ and $16$.
It can be proved easily that the Hadamard gate can be implemented
with arbitrary accuracy by choosing a proper $\kappa$.

\subsubsection{Arbitrary elements of SU(2)}
\label{subsec:arb1q}
\begin{figure}[htbp]
 \scalebox{0.90}[0.90]{\includegraphics{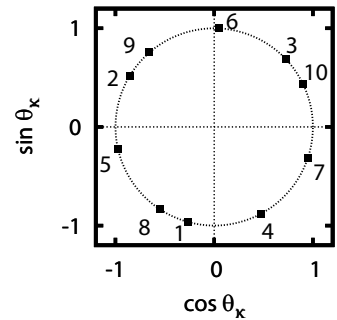}}
\caption{\label{fig:single_x} Rotation angles 
$\theta_{\kappa}=2\kappa \pi/\sqrt{3}$ in
 $\Gamma^{(1)}_{(1,\,0,\,0)}(\kappa)$
are plotted as points 
 $(\cos\theta_{\kappa}, \sin\theta_{\kappa})$ on the unit circle. 
The numbers in the figure indicate the values of $\kappa$. }
\end{figure}

The holonomy
$\Gamma^{(1)}_{(1,0,0)}(\kappa)$ 
is a quantum gate generating a rotation around the $x$-axis by an angle
$\theta_{\kappa}=\sqrt{2}\kappa \pi$. 
Although the set of $\theta_{\kappa}$ is discrete,  
we can find $\kappa$ which satisfies 
$|(\theta -\theta_{\kappa})\,\,{\rm mod}\,\, 2\pi|<\epsilon $ 
for arbitrary $\theta$ and $\epsilon$\,\cite{comment2}.
Figure \ref{fig:single_x} shows the points 
$(\cos\theta_{\kappa}, \sin\theta_{\kappa})$ 
for $\kappa =0, 1, \ldots, 10$. 
The point $(1,\,0)$ corresponds to $\kappa=0$ (i.e.,
$X=0$), in which no gate operation is performed. 
Similarly, we obtain an approximate rotation around the $y$-axis
with arbitrary accuracy by taking $\vm=(0,\,1,\,0)$.  
Consequently, it is possible to implement an arbitrary element of SU(2).

\section{Two-qubit gates}
\label{sec:hqg2}
\subsection{Hamiltonian}
\label{subsec:Hamiltonian2}
A logical two-qubit system consists of two dimers. 
The Hamiltonian is 
\begin{eqnarray}
 H_{\rm 2D}
&=&
H^{1} + H^{2},
\label{eq:Hamiltonian_for_twoq}
\end{eqnarray}
where 
\(
 H^{1}
=
-J_{1} \sigma_{1z} - J_{1}\,\sigma_{2z}
+ J_{1}\, \sigma_{1z}\sigma_{2z}
\) and 
\(
 H^{2}
=
-J_{2} \sigma_{3z} - J_{2}\,\sigma_{4z}
+ J_{2}\, \sigma_{3z}\sigma_{4z}
\). 
Equation (\ref{eq:Hamiltonian_for_twoq}) is used as $H$ in
Eq.~(\ref{eq:isodeform}). 
Here
$H^{1}$ ($H^{2}$) is the Hamiltonian of a single dimer
to which the first and the second (the third and the fourth) spins belong.
We easily find the $9$-fold degenerate lowest eigenvalue $-J_1-J_2$. 

We take the following coding space $\mathcal{C}_{2}$ 
for the logical two-qubit system: 
\(
 \mathcal{C}_{2}
= {\rm Span}\{
|T_{+}\rangle_1|T_{+}\rangle_2,\,|T_{+}\rangle_1|T_{0}\rangle_2,\,
|T_{0}\rangle_1|T_{+}\rangle_2,\,|T_{0}\rangle_1|T_{0}\rangle_2\}
\). 
The vector $|T_{+}\rangle_1$ denotes the eigenvector $|++\rangle$
associated with $H^{1}$, for example.

\subsection{Controlled-$e^{i\theta Z}$ gate}
\label{subsec:cz}
We choose the following generator of the isospectral 
deformation in Eq.~(\ref{eq:isodeform});
\begin{eqnarray}
 X
&=& X^{1} + X^{2} + X^{1\text{-}2},
\label{eq:X_2D}
\end{eqnarray}
where 
\(
 X^{1}
= i\vn_{1}\Omega_{1}\cdot (\vsigma_{1}+\vsigma_{2})
\), 
\(
 X^{2}
=
i\vn_{2}\Omega_{2}\cdot
(\vsigma_{3} + \vsigma_{4})
\) and 
\(
 X^{1\text{-}2}
= iJ (\sigma_{1\,z}\sigma_{3\,z} +\sigma_{1\,z}\sigma_{4\,z} 
+\sigma_{2\,z}\sigma_{3\,z} +\sigma_{2\,z}\sigma_{4\,z})
\). 
Here, $\vn_{1}$ and $\vn_{2}$ are unit vectors in
$\mathbb{R}^{3}$, while $\Omega_{1}$, $\Omega_{2}$, and $J$ are
positive real numbers. 
No undesired transitions into the
non-coding space take place for this choice of $X$. 
We elaborate this point in Appendix~\ref{appen:dexp_2X}. 

Let us find the corresponding anti-Hermitian connection with the
following unit vectors $\vn_{1}$ and $\vn_{2}$; 
\begin{eqnarray}
 \vn_{1}
&=&(0,\,0,\,1), \label{eq:vn12}\\
 \vn_{2}
&=&\left(\sqrt{1-n^{2}_{2z}},\,\,0,\,\,n_{2z}\right),
\label{eq:vn34} \\
 n_{2z}
&=& -\frac{J}{\Omega_{2}} \label{eq:vn34_z}.
\end{eqnarray}
We assume
\(
\Omega_{2} > J
\),
which guarantees $[H_{2D},\,X] \ne 0$.
Then, we obtain the following anti-Hermitian connection: 
\begin{eqnarray}
\left. A\right|_{\mathcal{C}_{2}}
&=&
 i\big[
\Omega_{1}I_{\rm L}\otimes I_{\rm L}
+
(\Omega_{1}+J)\sigma_{{\rm L}z}\otimes I_{\rm L} \nonumber \\
&+&\sqrt{2}\Omega_{2}n_{2x} I_{L}\otimes\sigma_{{\rm L}x}
+J\sigma_{{\rm L}z}\otimes \sigma_{{\rm L}z}
\big]. \label{eq:aconnection_code_controll}
\end{eqnarray}
Note here that
\(
\Omega_{2}n_{2x} = \sqrt{\Omega^{2}_{2}-J^{2}}
\).

The condition $e^{X}=\openone$ restricts the parameters as 
\begin{eqnarray}
 \Omega_{2}
&=& \kappa_{+}\pi \quad (\kappa_{+}\in\mathbb{N}),
\label{eq:cond_closed_34}\\
 \sqrt{\Omega_{2}^{2} + 8J^{2}}
&=& \kappa_{-}\pi
\quad (\kappa_{-}\in\mathbb{N}),
\label{eq:cond_closed_34p} \\
 \Omega_{1}
&=& \kappa^{\prime}\pi \quad (\kappa^{\prime}\in\mathbb{N}).
\label{eq:cond_closed_12}
\end{eqnarray}
Their derivation is given in Appendix~\ref{appen:dexp_2X}. 
It follows from Eqs.~(\ref{eq:cond_closed_34}) and (\ref{eq:cond_closed_34p})
that $ \kappa_{-} > \kappa_{+}$ and
$J=\frac{\pi}{2\sqrt{2}}\,\sqrt{\kappa_{-}^{2} - \kappa_{+}^{2}}$.
Note that the assumption $\Omega_{2}> J$
is equivalent to $ 3\kappa_{+} > \kappa_{-}$.

As a result, we obtain a two-qubit gate
\begin{eqnarray}
 \Gamma^{(2)}(\vkappa,\,\kappa^{\prime})
&=&
 (-1)^{\kappa^{\prime}}\,
\Gamma^{\rm LU}(\vkappa,\,\kappa^{\prime})\,\Gamma^{\rm C}(\vkappa), 
\end{eqnarray}
where 
\(
\vkappa = (\kappa_{+},\,\kappa_{-})
\). 
The local unitary gate $\Gamma^{\rm LU}(\vkappa,\,\kappa^{\prime})$ is
defined as 
\begin{equation}
\Gamma^{\rm LU}(\vkappa,\,\kappa^{\prime})
=
 e^{-i(\kappa^{\prime}\pi+J)\sigma_{{\rm L}z}}
\otimes
 e^{-i\nu\vk\cdot\vsigma_{\rm L}},
\end{equation}
where 
\(
 \nu = \sqrt{2\kappa^{2}_{1}\pi^{2}-J^{2}}
\) and
\(
 \nu\vk = (\sqrt{2\kappa^{2}_{1}\pi^{2}-2J^{2}},\,0,\,J)
\). 
The essential nonlocal operation is $\Gamma^{{\rm C}}(\vkappa)$, which is a
controlled-$e^{i \theta Z}$ gate and is given by
\begin{eqnarray}
\Gamma^{\rm C}(\vkappa)
=
 |0\rangle_{\rm L}\langle 0|_{\rm L}\otimes I_{\rm L}
+ |1\rangle_{\rm L}\langle 1|_{\rm L}\otimes e^{i2J\sigma_{{\rm L}z}}.
\end{eqnarray}
Thus, $\Gamma^{(2)}(\vkappa,\,\kappa^{\prime})$ followed by the
local unitary gate $\Gamma^{\rm LU}(\vkappa,\,\kappa^{\prime})^{\dagger}$
implements the controlled-$e^{i \theta Z}$ gate with $\theta = 2J$

The rotation angle $2J$ in $ \Gamma^{\rm C}(\vkappa)$ is characterized by 
$\kappa_{+}$ and $\kappa_{-}$.
We can, however, find $\kappa_{+}$ and $\kappa_{-}$ which satisfies 
$|(\theta - 2J)\,\,{\rm mod}\,\, 2\pi|<\epsilon$ 
for arbitrary $\theta$ and $\epsilon$\,\cite{comment3}. 
\begin{figure}[htbp]
\scalebox{1.0}[1.0]{\includegraphics{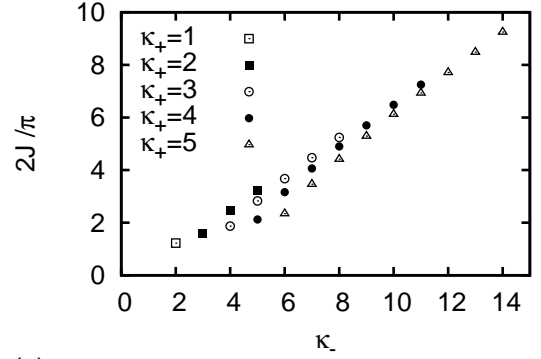}}
\scalebox{1.0}[1.0]{\includegraphics{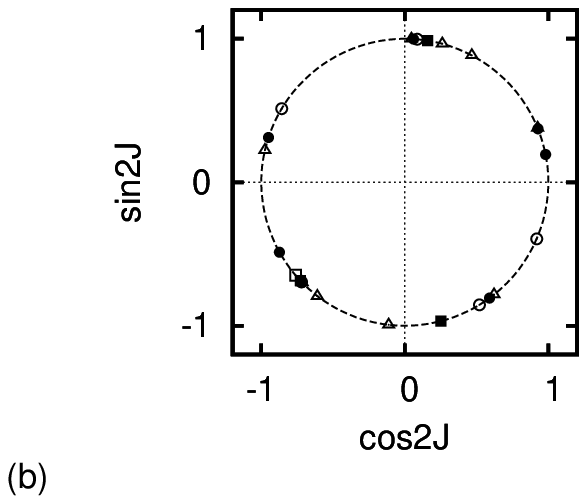}}
\caption{\label{fig:controll_z} Rotation angle $2J$ in the
 controlled-$e^{i\theta Z}$, $\Gamma^{C}(\bm{\kappa})$. (a) The value of $2J$
 with the unit of $\pi$ is plotted with respect to $\kappa_{-}$, when the fixed
 values of $\kappa_{+}$ are given. (b) The values of $\cos 2J$
 and $\sin 2J$ are plotted as the points on the unit circle. The
 plotting marks are the same meaning as in Fig.\,\ref{fig:controll_z}(a).}
\end{figure}
We show the attainable values of $2J$ for the various sets of
$(\kappa_{+},\,\kappa_{-})$ in Fig.~\ref{fig:controll_z}~(a). 
Note that $(\kappa_{+},\,\kappa_{-})$ have to be chosen such that
$\kappa_{+}<\kappa_{-}<3\kappa_{+}$ is satisfied. 
We calculate $(\cos 2J,\sin 2J)$ as in Fig.~\ref{fig:controll_z}~(b). 
Thus, a tunable coupling control scheme, in which $J$ is controllable, 
is necessary to achieve various
rotation angles in $\Gamma^{{\rm C}}(\bm{\kappa})$. 

Alternatively, by repeating the above gate $n$ times, we can implement the
controlled-$e^{i 2n J\sigma_{z}}$ gate. For an irrational $J$ and a given
$\theta$, 
it is possible to find $n$ such that 
$|(2n J - \theta)\ {\rm mod}\,2\pi| < \epsilon$ for an arbitrary
$\epsilon$. 
In this way, one may implement
the controlled-$e^{i \theta \sigma_z}$ gate with arbitrary precision.
In particular, we can implement the controlled-Z gate, which is a
constituent of the universal set of quantum gates, by taking 
$\theta = \pi/2$. 

\section{Summary}
\label{sec:summary}
We have constructed holonomic quantum gates
using isospectral deformations of an Ising model Hamiltonian.
These gates are the Hadamard gate, rotations around $x$- and $y$-axes 
and the controlled-$e^{i\theta Z}$ gate.
The closure of the loop in the parameter manifold 
leads to a discrete set of gates. 
These gates are, however, dense in
SU(2) and SU(4) and any one- and two-qubit gates 
can be implemented with arbitrary accuracy. 
A spin-chain model is a good candidate to implement holonomic quantum
gates with the ground state eigenspace.  
We will propose a more feasible scheme based on the present
analyses in our future work.    

\begin{acknowledgments}
This work was supported by ``Open Research Center'' Project for
 Private Universities: matching fund subsidy from MEXT (Ministry of
 Education, Culture, Sports, Science and Technology). 
MN's work is partially supported by the Grant-in-Aid for Scientific Research
(C) from JSPS (Grant No. 19540422). 
YO would like to thank to Shuzo Izumi, Toshiyuki Kikuta and Kaori Minami for
 valuable discussions. 
\end{acknowledgments}

\appendix
\section{Emulation of holonomic quantum gates}
\label{appendix:exact}
An exact solution of the Schr\"odinger equation is obtained
when the Hamiltonian is deformed according to Eq.\,(\ref{eq:isodeform}).
Let us consider the Schr\"odinger equation in term of the dimensionless
time $\tau$:
\begin{equation}
 i\frac{d}{d\tau}|\psi(\tau)\rangle = TH(\tau)|\psi(\tau)\rangle,
\label{eq:sch_eq}
\end{equation}
where 
\(
H(\tau) = e^{X\tau}\,H\,e^{-X\tau}
\). 
Introducing $|\phi(\tau)\rangle = e^{-X\tau}|\psi(\tau)\rangle$, we obtain the
following equation for $|\phi(\tau)\rangle$:
\begin{equation}
 i\frac{d}{d\tau}|\phi(\tau)\rangle 
= \left(
-iX + H
\right)|\phi(\tau)\rangle.
\label{eq:int_sch_eq}
\end{equation}
Accordingly, the time evolution operator at $\tau=1$ is 
\begin{equation}
 U(\tau=1) = e^{X}\,e^{-i(-iX + HT)}.
\label{eq:pulse_eq} 
\end{equation}
When the adiabatic approximation is valid and the
initial state is in the lowest eigenspace, we obtain
\begin{equation}
 U(\tau=1)P_{0} = e^{-iE_{0}T}\Gamma P_{0}, 
\label{eq:ad_eq} 
\end{equation}
where $P_{0}$ is the projection operator on the lowest eigenspace.  
From Eqs.~(\ref{eq:pulse_eq}) and (\ref{eq:ad_eq}), we obtain
\begin{equation}
 e^{X}\,e^{-i(-iX + HT)}P_{0} 
= 
e^{-iE_{0}T}\,
\Gamma\,P_{0}.
\label{eq:simulation_hqg}
\end{equation}
The right hand side of Eq.~(\ref{eq:simulation_hqg}) amounts to a holonomic
quantum gates, up to the dynamical phase $e^{-iE_{0}T}$.
The left hand side of Eq.~(\ref{eq:simulation_hqg}) may be interpreted as
implementation of the quantum dynamics by a pulse sequence.
Equation.~({\ref{eq:simulation_hqg}}) provides a way how to emulate 
a holonomic gate $\Gamma$ with a series of pulses of $e^{-i(-iX+HT)}$ and
$e^{X}$, although it may not be a genuine realization.

\section{Generator of isospectral deformation for two-qubit gates} 
\label{appen:dexp_2X}
Let us analyze the the generator of the isospectral deformation for
two-qubit gates in Sec.~\ref{subsec:cz}.
In this Appendix, we denote
the eigenvector $|T_{+}\rangle_{1}|T_{+}\rangle_{2}$ simply as 
$|T_{+}T_{+}\rangle$, for example.

First, we focus on $X^{1}$ and $X^{2}$ in Eq.~(\ref{eq:X_2D}). 
We obtain
\begin{widetext}
\begin{eqnarray}
 X^{1} = i\Omega_{1}\left(
\begin{array}{ccc}
2n_{1z} & 
\sqrt{2}(n_{1x}-in_{1y}) & 
 0\\
\sqrt{2}(n_{1x}+in_{1y}) &
0 & 0 \\
0 &
0 &  0
\end{array}
\right), \quad
 X^{2} =  
i\Omega_{2}\left(
\begin{array}{ccc}
2n_{2z} & 
\sqrt{2}(n_{2x}-in_{2y}) & 
 0\\
\sqrt{2}(n_{2x}+in_{2y}) &
0 & 0 \\
0 &
0 &  0
\end{array}
\right). \label{eq:mform_X_1_2}
\end{eqnarray} 
\end{widetext}
The matrix element $(X^{i})_{kl}$ corresponds to 
\(
_{i}\langle k|X^{i}|l \rangle_{i}
\), where 
\(
|1\rangle_{i} = |T_{+}\rangle_{i}
\), 
\(
|2\rangle_{i} = |T_{0}\rangle_{i}
\), and 
\(
|3\rangle_{i} = |S_{0}\rangle_{i}
\) $(i=1,\,2)$. 
One can easily find that 
\(
\langle T_{+}S_{0}|X^{1}|T_{+}T_{0}\rangle =0
\), for example. 
This means that undesired transitions into the subspace irrelevant to quantum
computation never occur under $X^{1}$ and $X^{2}$. 

Next, let us consider $X^{1\text{-}2}$ in Eq.~(\ref{eq:X_2D}). 
The following observations are useful to calculate the matrix
elements: 
\begin{eqnarray}
 \sigma_{kz}|T_{+}\rangle_{i} &=& |T_{+}\rangle_{i}, 
\label{eq:baseid1}\\
 \sigma_{kz}|T_{0}\rangle_{i} &=& (-1)^{k+1}|S_{0}\rangle_{i}, 
\label{eq:baseid2} 
\end{eqnarray}
where $k=1,\,2$ for $i=1$ and $k=3,\,4$ for $i=2$. 

Thus, we only have to consider five elements 
\(
\langle T_{+}T_{+}|X^{1\text{-}2}|T_{+}T_{+}\rangle
\), 
\(
\langle T_{+}S_{0}|X^{1\text{-}2}|T_{+}T_{0}\rangle
\), 
\(
\langle S_{0}T_{+}|X^{1\text{-}2}|T_{0}T_{+}\rangle
\), 
\(
\langle S_{0}S_{0}|X^{1\text{-}2}|T_{0}S_{0}\rangle
\), and 
\(
\langle S_{0}T_{0}|X^{1\text{-}2}|T_{0}S_{0}\rangle
\). 
Note that one can easily calculate the other non-trivial elements (e.g., 
\(
\langle T_{0}T_{+}|X^{1\text{-}2}|S_{0}T_{+}\rangle
\)) from the anti-Hermitian property of $X^{1\text{-}2}$. 
The other elements which are related to the lowest eigenspace of
$H_{2D}$ trivially vanish.
As a result, we obtain the following results:
\(
\langle T_{+}T_{+}|X^{1\text{-}2}|T_{+}T_{+}\rangle 
= i4J
\) and 
\(
\langle T_{+}S_{0}|X^{1\text{-}2}|T_{+}T_{0}\rangle
=
\langle S_{0}T_{+}|X^{1\text{-}2}|T_{0}T_{+}\rangle
=
\langle S_{0}S_{0}|X^{1\text{-}2}|T_{0}S_{0}\rangle
=
\langle S_{0}T_{0}|X^{1\text{-}2}|T_{0}S_{0}\rangle
=0
\). 
As a result, all the matrix elements corresponding to the undesired transition
to the irrelevant subspace vanish.

Finally, we calculate $e^{X}$ when the unit
vectors $\vn_{1}$ and $\vn_{2}$ are given by Eqs.\,(\ref{eq:vn12})
and (\ref{eq:vn34}), respectively. 
We obtain 
\begin{eqnarray}
 e^{X}
&=&
 |T_{+}\rangle_{11}\langle T_{+}|\otimes e^{iY_{+}}
+ |T_{-}\rangle_{11}\langle T_{-}|\otimes e^{iY_{-}} 
 \nonumber \\
&&
\quad + 
(|T_{0}\rangle_{11}\langle T_{0}|
+|S_{0}\rangle_{11}\langle S_{0}|)
\otimes e^{iY_{0}}, \\
 Y_{0}
&=&
 \Omega_{2}\vn_{2}\cdot (\vsigma_{3}+\vsigma_{4}),\\
 Y_{\pm} 
&=& \pm 2\Omega_{1}\openone_{2} +  \nu_{\pm}\vk_{\pm}
\cdot (\vsigma_{3} + \vsigma_{4}),
\end{eqnarray}
where
\(
\nu_{+} = \Omega_{2}
\), 
\(
\nu_{-} = \sqrt{\Omega^{2}_{2}+8J^{2}}
\), and 
\(
\openone_{2} = 
|T_{+}\rangle_{22}\langle T_{+}| + 
|T_{0}\rangle_{22}\langle T_{0}| + 
|T_{-}\rangle_{22}\langle T_{-}| + 
|S_{0}\rangle_{22}\langle S_{0}| 
\). 
In particular, the values of $\nu_{\pm}$ is important for imposing the
condition $e^{X}=\openone$. 
The unit vector $\vk_{\pm}$ is given by
\(
\nu_{\pm}\vk_{\pm}
=
(\Omega_{2}n_{2x},\,0,\,\Omega_{2}n_{2z}\pm 2J)
\).
The condition $e^{X}=\openone$ implies 
\(
e^{iY_{0}}=\openone_{2}
\) and 
\(
e^{iY_{\pm}} = \openone_{2}
\). 
The condition $e^{iY_{0}}=\openone_{2}$ implies there is
$\kappa_{+}\in\mathbb{N}$ such that 
\(
\Omega_{2} = \kappa_{+}\pi
\); this requirement is equivalent to Eq.~(\ref{eq:cond_closed_34}). 
Similarly, if Eqs.~(\ref{eq:cond_closed_34p}) and
(\ref{eq:cond_closed_12}) are satisfied, then we obtain
$e^{iY_{\pm}}=\openone_{2}$.


\begin{thebibliography}{99}
\bibitem{comment}
It seems that the condition 
$H(1) = H(0)=H$ is not properly taken into 
account in Ref.~\cite{KarimipourMajd2005}. 
They take 
\(
H = 2 J (\sigma_{1z} + \sigma_{2z}) + J{\bm \sigma_1}\cdot {\bm
	\sigma_2} 
\), 
for a one-qubit gate and 
\(
X = i \frac{\pi}{2}(\sigma_{1x}-\sqrt{2}\sigma_{1z})
\), 
for the Hadamard gate.
Consequently, the coefficient of $\sigma_{1z}$ in $H(1)$ is 
\(
 2J\cos\sqrt{3} + \frac{8J}{3}\sin^{2}\frac{\sqrt{3}}{{2}}
\), 
which shows $H(1) \neq H(0)$.
\bibitem{comment2}
If an irrational number $\alpha$ is given, the set
$\{(n\alpha)|\,n\in\mathbb{N} \}$ is dense in
	 the interval $(0,\,1)$, where
	 $(n\alpha)=n\alpha-\lfloor n\alpha \rfloor$ and 
$\lfloor n\alpha \rfloor$ is the greatest integer that is less than or equal to
	 $n\alpha$\,\cite{HardyWright1978}. 
Accordingly, for arbitrary $\theta$ and $\epsilon>0$, there is
	 $n\in\mathbb{N}$ such that $|e^{in\alpha\pi} -
	 e^{i\theta}|<\epsilon$.
\bibitem{comment3}
Let us first put $\kappa_{+}=u$ and $\kappa_{-}=v$ ($u<v<3u$), where the
value of $\frac{1}{\sqrt{2}}\sqrt{v^{2}-u^{2}}(\equiv \gamma)$ is an
irrational number; for example, $\gamma=\sqrt{\frac{3}{2}}$ when
$(u,\,v)=(1,2)$. 
Next, we redefine the values of $\kappa_{+}$ and $\kappa_{-}$ as follows:
$\kappa_{+}=lu$ and $\kappa_{-}=lv$, where $l\in\mathbb{N}$.
Note that those values satisfy $\kappa_{-}>\kappa_{+}$ and
$3\kappa_{+}>\kappa_{-}$, because $u<v<3v$ and $l>0$.
Thus, the rotation angle $2J$ is \(l\gamma\pi\). 
Therefore, we can find a proper $l\in\mathbb{N}$ such that approximates
	 a given rotation angle with arbitrary accuracy\,\cite{comment2}.
\end{thebibliography}
\end{document}